\documentclass[twocolumn,showpacs,preprintnumbers,amsmath,amssymb]{revtex4}

\usepackage{graphicx}
\usepackage{dcolumn}
\usepackage{bm}

\begin{document}      

\title{The high-field phase diagram of the cuprates derived from the Nernst effect} 
\author{Yayu Wang$^1$, N. P. Ong$^1$, Z. A. Xu$^{1,\dagger}$, T. Kakeshita$^2$, S. Uchida$^2$, 
D. A. Bonn$^3$, R. Liang$^3$ and W. N. Hardy$^3$}      
\address{
$^1$Department of Physics, Princeton University, Princeton, New Jersey 08544. 
}
\address{$^2$School of Frontier Sciences, University of Tokyo, Bunkyo-ku, Tokyo 113-8656, 
Japan.}
\address{$^3$Department of Physics and Astronomy, University of British Columbia, Vancouver, BC V6T 
1Z1, Canada.}
\date{\today}      

\begin{abstract}
Measurements of the Nernst signal in the vortex-liquid state of the cuprates to high fields (33 T) reveal that 
vorticity extends to very high fields even close to the zero-field critical temperature $T_{c0}$.  In 
overdoped $\rm La_{2-x}Sr_xCuO_4$ (LSCO) we show that the upper critical field $H_{c2}(T)$ curve 
does not end at $T_{c0}$, but at a much higher temperature.   These results imply that $T_{c0}$ 
corresponds to a loss in phase rigidity rather than a vanishing of the pairing amplitude.  An intermediate field 
$H^*(T)\ll H_{c2}(T)$ is shown to be the field scale for the flux-flow resistivity.
\end{abstract}
\maketitle                   

In the cuprate superconductors, strong fluctuations of the order parameter~\cite{Fisher,Blatter} and the 
extreme field scales make the task of constructing the experimental phase diagram a formidable challenge.  
While the important vortex-solid melting line $H_m(T)$ is now well-known~\cite{Zeldov,Schilling,Liang}, 
there is great uncertainty in the region above $H_m$.  A key difficulty is the {\em loss of long-range phase 
rigidity} in the transition to the vortex-liquid state.  In the vortex-liquid state, the flux-flow resistivity $\rho$ 
rises very rapidly to the (extrapolated) normal-state value~\cite{Chien} even though substantial condensate 
strength and pairing amplitude remain.  This makes $\rho$ an unreliable `diagnostic' of the field-suppression 
of the pairing amplitude.  By contrast, the Nernst and Ettinghausen effects can probe the existence of 
vorticity in the superfluid with undiminished sensitivity even in intense fields.  From Nernst measurements in  
$\rm La_{2-x}Sr_xCuO_4$ (LSCO) and $\rm YBa_2Cu_3O_y$ (YBCO) in fields up to 33 T, we have 
uncovered an anomalous property of $H_{c2}(T)$ near their zero-field critical temperature $T_{c0}$.  
This anomaly is related to the existence of vortex-like excitations high above $T_{c0}$~\cite{Xu,Wang}, 
and is central to the key question of whether the Meissner transition in zero field is caused by the collapse of 
long-range phase coherence or the vanishing of the pairing amplitude (see also Corson {\it et 
al.}\cite{Corson}).  In YBCO, previous Ettinghausen~\cite{Palstra} and Nernst ~\cite{Ri,Clayhold} 
measurements were performed in lower fields.  Measurements on LSCO have been extended recently to 
high fields~\cite{Capan}.

\begin{figure}				
\includegraphics[width=9cm]{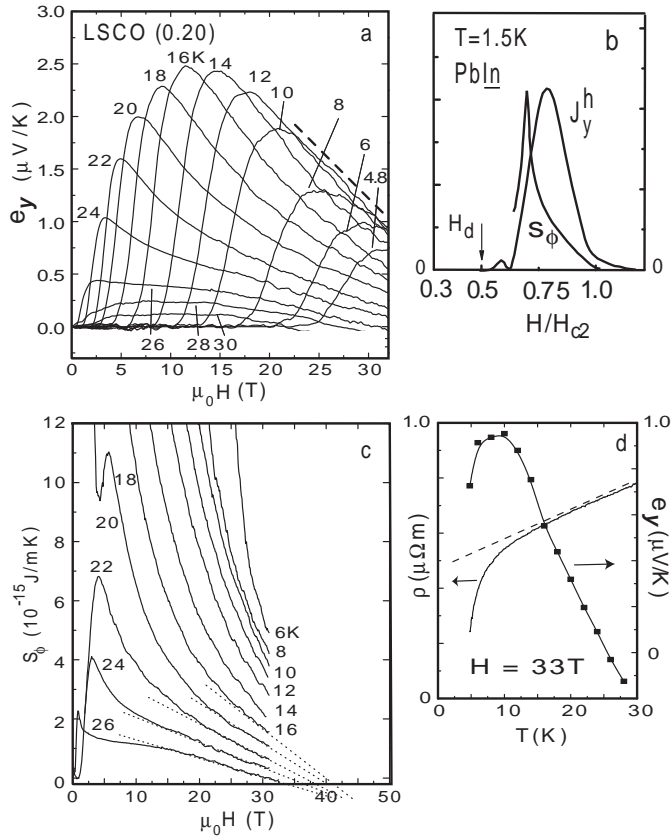}
\caption{\label{L20} (al) The field dependence of $e_y\equiv E_y/|\nabla T|$ at indicated $T$ in Sample 1 
(LSCO with $x$ = 0.20, $T_{c0}$ = 28 K).  Below 25 K, $e_y$ rises to a sharp peak before decreasing 
montonically.  Broken line indicates asymptotic behavior at large-$H$ and low $T$.  (b) Plots of the 
transverse heat flux $J^h_y$ and $s_{\phi}$ vs. $H$ in an Ettinghausen experiment in Pb\underline{In} at 
1.5 K (by reciprocity, $J^h_y = e_y/JT$, with $J$ the applied supercurrent).  $H_d$ is the depinning field 
(modified from Vidal \protect\cite{Vidal}a).  (c) The field dependence of the transport line-entropy 
$s_{\phi}$ obtained from $e_y$ and $\rho$ in Sample 1 (LSCO, $x$ = 0.20).  The high-field linear 
extrapolations (broken lines) provide estimates of $H_{c2}(T)$.  Fits to Eq. \ref{Maki} give $\kappa$ = 
670 and 500 at 26 and 18 K, respectively.  (d) The temperature dependence of $\rho$ and $e_y$ in 
Sample 1 in a fixed field (33 T).  The broken line is the extrapolation of $\rho_N$.  Note that $e_y$ 
(measured at 33 T) goes to zero at $\sim$26 K instead of $\sim$9 K (the `knee' in $\rho$).
}
\end{figure}

In the Nernst experiment, vortices moving with velocity ${\bf v}$ down a thermal gradient $-\nabla 
T\parallel {\bf\hat{x}}$ generate a Josephson voltage which is observed as a transverse $E$-field $E_y = 
Bv_x$, with $\bf B$ the mean flux density.  The vortex-Nernst effect is well-explored in low-$T_c$ 
superconductors.  Figure \ref{L20}b shows the Ettinghausen-Nernst effect in 
Pb\underline{In}~\cite{Vidal}.  The depinning of the vortex lattice by a supercurrent ($\bf J\parallel 
\hat{x}$) leads to a large transverse heat current $J^h_y\parallel {\bf \hat{y}}$ that rises to a maximum and 
then decreases to zero at $H_{c2}$ (the weak fluctuation tail above $H_{c2}$ is not relevant here).  By 
Onsager reciprocity, the Nernst signal $e_y\equiv E_y/|\nabla T|$ shares the same profile as $J^h_y$ in Fig. 
\ref{L20}b~\cite{Onsager}.  To compare with theory, $e_y$ is divided by $\rho(H)$ to obtain the 
`transport line-entropy' $s_{\phi} = \phi_0 e_y/\rho$ ($Ts_{\phi}$ is the heat energy carried by a 
unit-length vortex line).  In Pb\underline{In}, the linear decrease in $s_{\phi}$ near $H_{c2}$ matches that 
of the magnetization $M(T,H) = -[H_{c2}(T)-H]/1.16[2\kappa^2-1]$ ($\kappa$ is the Ginzburg-Landau 
parameter), as given by the Caroli-Maki expression \cite{Maki,Vidal}
\begin{equation}
s_{\phi}(T,H) = \frac{\phi_0}{T} |M(T,H)| L_D(T) \label{Maki}
\end{equation}
(here $L_D(T)$ decreases gradually with $T$ from 1 at $T_{c0}$).

We start with the overdoped regime, where the field profiles of $e_y$ most resemble those in 
Pb\underline{In}. In Fig. \ref{L20}a, we display curves of $e_y$ vs. field $H$ in Sample 1 [$\rm 
La_{2-x}Sr_xCuO_4$ (LSCO) with $x$ = 0.20 and $T_{c0}$ = 28 K] (for experimental details see Ref. 
\cite{Wang}).  At each temperature $T$, the signal rises steeply above $H_m$, attaining a prominent 
maximum before decreasing.  The total data set defines experimentally the region in $H$ and $T$ where 
vorticity is strongly present.  As in Pb\underline{In}, the monotonic decrease at high fields reflects the field 
suppression of the condensate strength.  At high fields, all the curves below 14 K are observed to follow a 
common curve towards zero (broken line).  The trend implies that the low-$T$ traces are all wedged 
between the field axis and the broken line.  Hence all the low-$T$ curves vanish at the intercept of the 
common curve with the field axis (45-50 T), which corresponds to $H_{c2}(0)$ .

\begin{figure}				
\includegraphics[width=8cm]{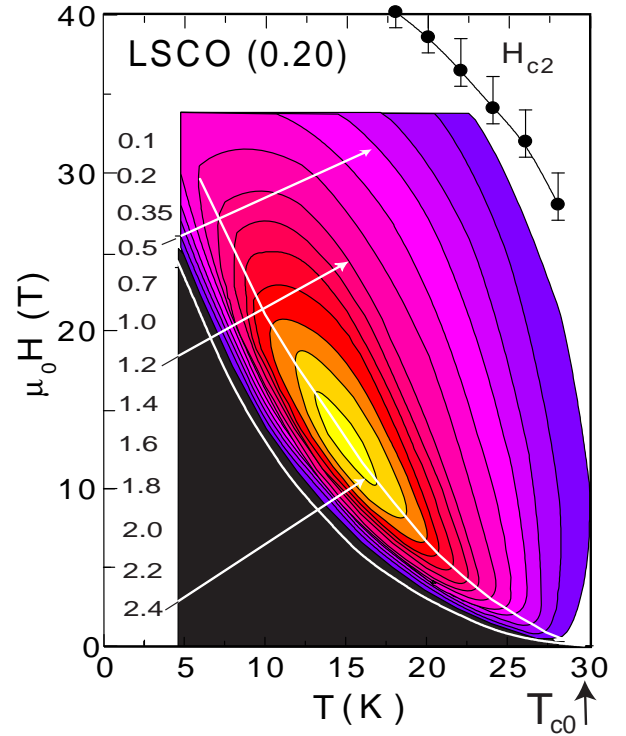}
\caption{\label{Lcont} The measured values of $e_y(T,H)$ in Sample 1 (LSCO, $x$ = 0.20) represented 
as a contour plot in the $T$-$H$ plane.  Values of $e_y$ (in $\mu$V/K) at successive contour lines are 
shown on the left column (white arrows).  The lower and upper white lines are the melting field $H_m$ and 
`ridge' field $H^*$, respectively.  $H_{c2}(T)$ values estimated from Fig. \ref{L20}c are shown as solid 
circles.  $T_{c0}$ is indicated in the lower-right corner.  
}
\end{figure}

Going to higher $T$, we immediately encounter an anomaly.  Conventionally, the $H_{c2}$ line goes 
linearly to zero at $T_{c0}$.  Hence $e_y$ ought to be finite in a field interval that $\rightarrow$ 0 as 
$T\rightarrow T_{c0}^-$.  In sharp contrast, we find that, close to $T_{c0}$, the magnitude of $e_y$ 
remains large and nearly unchanged up to intense fields.  There is no evidence for a field scale (flagged by 
$e_y\rightarrow 0$) that decreases to zero regardless of how close we are to $T_{c0}$.  To make this 
point quantitative, we convert $e_y$ to $s_{\phi}$ using $\rho(T,H)$ measured in the same sample.  The 
line entropy rises steeply to a sharp maximum before decreasing monotonically (Fig. \ref{L20}c).  The linear 
decrease at high fields, strikingly similar to that in Pb\underline{In}, allows $H_{c2}(T)$ to be 
determined~\cite{fit}.  These results show that $H_{c2}(T)$ decreases with rising $T$, {\em but remains at 
extraordinarily high values} as $T\rightarrow T_{c0}$ (see Fig. \ref{Lcont}).  [We neglect a small 
correction from the negative Nernst coefficient $\nu_N\simeq$ -40 nV/KT of the holes in overdoped LSCO 
(Fig. 3 of Ref. \cite{Xu}).  Accounting for the background increases $s_{\phi}$ and $H_{c2}(T)$ by $\sim 
10\%$. ]

A powerful way to summarize the anomalous trends described is to represent $e_y(T,H)$ as a contour plot 
in the $T$-$H$ plane (Fig. \ref{Lcont}).  In the vortex-solid phase below the melting line $H_m$ (lower 
white curve), $e_y=0$ because the vortices are pinned.  As the vortex solid melts across $H_m$, $e_y$ 
increases steeply up to the `ridge' which represents a field scale that we call $H^*(T)$ (upper white curve).  
To the right of the ridge, $e_y$ falls monotonically.  We focus on the two interesting regimes: the low-$T$ 
limit and the region near $T_{c0}$.  

As $T\rightarrow 0$, the two fields $H_m(T)$ and $H^*(T)$ increase rapidly towards values approaching 
$H_{c2}(0)$~\cite{melting}.   In this large-$H$, low-$T$ region, the contours are nearly horizontal, as 
required by the asymptotic high-field behavior (broken curve in Fig. \ref{L20}a).  This implies that the 
$H_{c2}$ line is nearly horizontal below $\sim$12 K, as anticipated above.  At temperatures $T\le 
T_{c0}$, the weak attenuation of $e_y$ to high fields corresponds to the nearly vertical orientation of the 
contours.  Clearly, there is no evidence for a putative $H_{c2}$ line that terminates at $T_{c0}$.  The 
values of $H_{c2}(T)$ derived from $s_{\phi}$ are plotted in Fig. \ref{Lcont} as solid 
circles.  

\begin{figure}				
\includegraphics[width=7cm]{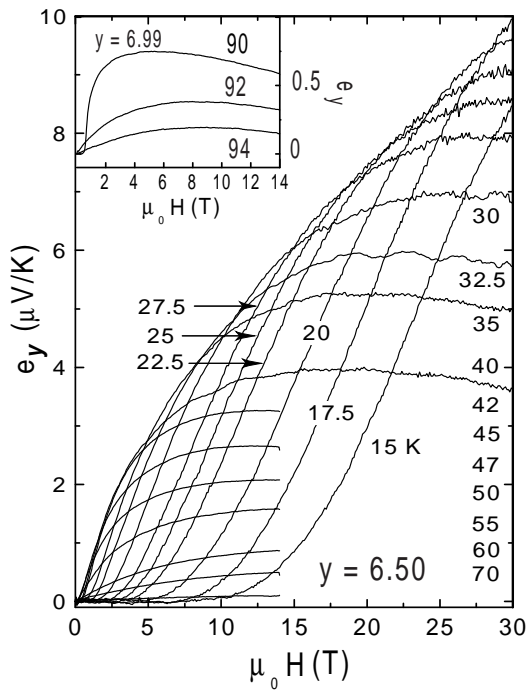}
\caption{\label{Y650} Variation of the Nernst signal $e_y\equiv E_y/|\nabla T|$ with field $H$ at fixed $T$ 
in Sample 2 (YBCO with $y$ = 6.50, $T_{c0}$ = 50 K).  Above 40 K, the measurements were 
performed up to 14 T only.  The insert shows $e_y$ vs. $H$ in Sample 3 (YBCO, $y$ = 6.99) close to its 
$T_{c0}$ = 93 K. Both are detwinned crystals grown in barium zirconate crucibles.
}
\end{figure}

The anomalous features of the Nernst signal become more pronounced when we go to the underdoped 
regime.  A major change from the overdoped case is apparent in the field profiles of $e_y$.  We show in 
Fig. \ref{Y650} results in underdoped YBCO (Sample 2, with $y = 6.50$ and $T_{c0}$ = 50 K).  As 
$H$ increases above $H_m$, $e_y$ rises rapidly, but attains a very broad maximum that extends 
undiminished to 30 T.  In comparison with the curves in Fig. \ref{L20}a, we see that $e_y$ in Sample 2 
approaches zero only at fields considerably higher than 30 T.  If we now convert $e_y$ to $s_{\phi}$ using 
the measured $\rho$, the broad profile translates to a line entropy that extrapolates to zero at fields much 
higher than our maximum field.  At 40 K, we may estimate the lower bound $H_{c2}(T) \ge$60 T, which is 
very high for a temperature so close to $T_{c0}$ = 50 K.  These features are common to all the 
underdoped cuprates we have investigated to date (for results in LSCO and Bi 2201, see Ref. 
\cite{Wang}).

\begin{figure}			
\includegraphics[width=8cm]{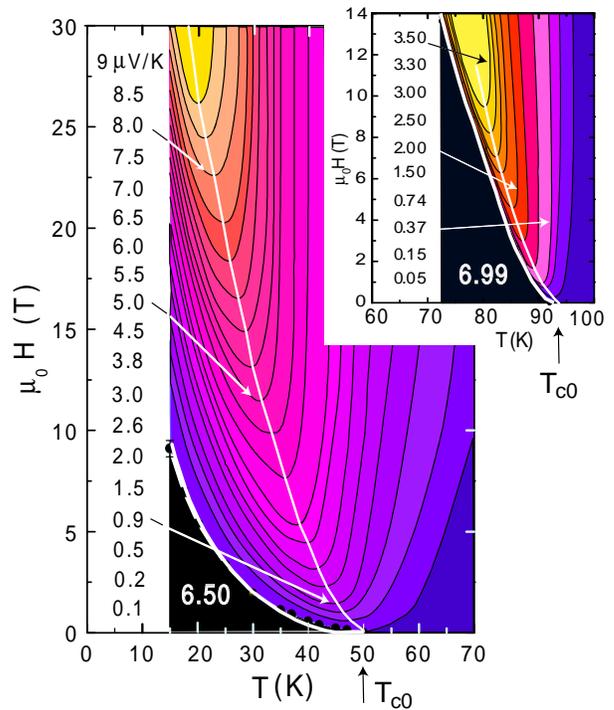}
\caption{\label{Ycont} [Main Panel] Contour plot of $e_y$ in the $T$-$H$ plane in Sample 2 (YBCO, 
$y$ = 6.50).   The upper (lower) white curve represents the `ridge' field $H^*$ (melting field $H_m$).    
Insert shows the corresponding contour plot in Sample 3 ($y$ = 6.99).  Note that the region between 
$H_m$ and $H^*$ is compressed to a thin sliver.  
}
\end{figure}

The contour plot of $e_y$ for Sample 2 is displayed in the main panel Fig. \ref{Ycont}.  In comparison with 
Fig. \ref{Lcont}, the melting line is now considerably lower relative to $H^*$.  Further, the contour lines in 
the region around $T_{c0}$ are more spread out.  The Nernst signal retains considerable strength at 70 K, 
indicating that vorticity strongly persists up to 20 K above $T_{c0}$.  The contour plots show rather clearly 
the {\em continuity} between the high-temperature fluctuation phase observed by Xu {\em et 
al.}~\cite{Xu,Wang} and the vortex-liquid state below $T_{c0}$.  The 2 regimes smoothly merge together, 
and no phase boundary or `crossover' line separating them is apparent in the $T$-$H$ plane.  As mentioned 
above, the upper critical field is estimated to be above 60 T even at 40 K.  The much higher field scale in 
underdoped YBCO is also apparent if we compare the contour plots in Figs. \ref{Lcont} and \ref{Ycont}.  
In overdoped LSCO, the contours close at fields above 13 T (the field scale of the `island') whereas, in 
underdoped YBCO, they do not close until $H$ exceeds 30 T.

For comparison, we also display the contours in near-optimal YBCO (Sample 3 with $y$ = 6.99 and 
$T_{c0}$ = 93 K).  As shown by the vertical orientation of the contour lines from 85 to 95 K, as well as 
the field profiles in the insert in Fig. \ref{Y650}, $e_y$ is virtually field independent up to 14 T.  Even for 
near-optimal YBCO, there is no evidence for an $H_{c2}$ line that decreases to zero at $T_{c0}$ (this 
calls into question the interpretation of `$H_{c2}$' values inferred from bulk magnetization 
results~\cite{Welp}).

The observation of vortex-like excitations high above $T_{c0}$ by Xu {\em et al.} ~\cite{Xu,Wang} raised 
the question whether the cuprate superconducting transition at $T_{c0}$ actually corresponds to the loss of 
long-range phase rigidity, as opposed to the vanishing of the Gorkov pairing amplitude ${\cal F}({\bf 
x,x}')$~\cite{Emery}.   The present results provide strong support for the former.  A prominent anomaly at 
temperatures near $T_{c0}$ is the remarkably weak field dependence of $e_y$ above the field scale 
$H^*(T)$, which implies that the condensate strength extends undiminished to very high fields.  In 
underdoped cuprates, this anomalous behavior persists to our lowest temperatures (Fig. \ref{Y650}).  
However, in overdoped samples, the profiles sharpen up below 25 K to reveal pronounced field 
suppression at high fields (Fig. \ref{L20}a).  Values of $H_{c2}(T)$ inferred from the line-entropy profiles 
show that the reason for the anomaly is that the $H_{c2}$ vs. $T$ line does not terminate at $T_{c0}$, but 
at a much higher temperature.   

This seems to us to be compelling evidence that, at $T_{c0}$, the steep increase of $\rho$ and the collapse 
of the Meissner effect, correspond to a pre-emptive loss of phase rigidity (as has been advocated by Emery 
and Kivelson \cite{Emery} and others) rather than the vanishing of ${\cal F}({\bf x,x}')$.  This scenario is 
qualitatively distinct from that in bulk low-$T_c$ superconductors.  Although $\rho$ saturates rapidly to 
$\rho_N$ above $T_{c0}$ (in zero field), there remains sufficient condensate density to support vorticity.  
In Sample 1, the measured $H_{c2}$ curve implies that we need to apply fields above 30 T to suppress all 
vorticity near $T_{c0}$.  The field scale is even higher in underdoped YBCO.  

A hallmark of the vortex-liquid state in cuprates is that $\rho$ rapidly increases and `saturates' to the 
extrapolated normal-state value ($\rho_N$) even when $H\ll H_{c2}(T)$. To illustrate this, we plot the $T$ 
dependence of $\rho$ and $e_y$ in Sample 1 with $H$ fixed at 33 T in Fig. \ref{L20}d (insert).  Clearly, 
the large Nernst signal indicates that the region from 5 to 25 K is in the vortex-liquid state (see Fig. 
\ref{Lcont}).  However, between 15 and 30 K, $\rho$ virtually lies on the extrapolated $\rho_N$ curve.  
Hence reliance on $\rho$ alone would lead to a serious underestimate of `$H_{c2}$'.  

In all cuprates studied, the ridge field $H^*(T)$ line is ubiquitous.  Over a broad range of $T$, the ratio 
$\rho/\rho_N$ has the value 0.3-0.4 on the curve $H^*$ vs. $T$.   $H^*(T)$ serves as the boundary 
separating the low-field regime in which $\rho$ rapidly increases with $H$, from the high-field regime in 
which $\rho$ slowly asymptotes to the extrapolated $\rho_N$.   Hence $H^*$ represents an intrinsic field 
scale that controls dissipation in the vortex liquid state (attempts to find $H_{c2}$ using $\rho$ alone 
invariably turn up a curve akin to $H^*$ rather than the higher $H_{c2}$).  If we identify a length scale 
$\xi^*(T)$ by the ratio $\xi^*(T)/\xi_0(T) \sim \sqrt{H_{c2}(T)/H^*(T)}$, we find that at low $T$, 
$\xi^*/\xi_0\approx 1$.  As $T$ increases, however, the ratio rapidly increases (to 1.8 and 3.7 at 15 and 
25 K, respectively).  Since $H^*$ determines the onset of strong flux-flow dissipation, $\xi^*$ represents a 
radius larger than the vortex core radius $\xi_0$ ($\simeq$2.6 nm).  Packing vortices closer than the outer 
radius $\xi^*$ leads to a crossover to the high-dissipation state in which $\rho$ asymptotes to $\rho_N$.  
However, as superfluidity and its associated 2$\pi$ phase winding survive right up to the inner radius $\xi_0$ 
\cite{Wen}, the Nernst signal is observable to the higher field scale $H_{c2}$ ($s_{\phi}$ is strongly 
attenuated in magnitude because of the reduced supercurrent for $\xi^*<r<\xi_0$)~\cite{fit}. A length scale 
extending outside the vortex core has been observed in tunneling experiments \cite{Pan}.

We thank S. Hannahs for valuable assistance at the National High Magnetic Field Lab. (a facility supported 
by NSF and the state of Florida).  We acknowledge useful discussions with P.W. Anderson, A. J. Millis and 
Z. Y. Weng.  This research is supported by a NSF grant (NSF-DMR 98-09483), the Office of Naval 
Research (N00014-01-0281), the New Energy and Industrial Tech. Develop. Org. (NEDO), Japan, the 
Natural Sciences and Engineering Research Council, and the Canadian Institute for Advanced 
Research.

\noindent
$^\dagger$ {\em Present address of ZAX: Department of Physics, Zhejiang University, Hangzhou, 
China}


\begin{references}
\bibitem{Fisher} D. S. Fisher, M. P. A. Fisher, and D. A. Huse, Phys. Rev. B. {\bf 43}, 130
    (1991). 
\bibitem{Blatter} G. Blatter, M. V. Feigel'man, V. B. Genshkenbein, A. I. Larkin, and V. M. Vinokur, Rev. 
Mod. Phys. {\bf 66}, 1125 (1994).
\bibitem{Zeldov} E. Zeldov, D. Majer, M. Konczykowski, V. B. Geshkenbein, V. M. Vinokur, and H. 
Shtrikman, Nature {\bf 375}, 373 (1995).
\bibitem{Schilling} A. Schilling, R. A. Fisher, N. E. Phillips, U. Welp, D. Dasgupta, W. K. 
Kwok, 
    and G. W. Crabtree, Nature {\bf 382},  791 (1996).
\bibitem{Liang}  R. Liang, D. A. Bonn, W. N. and Hardy, Phys. Rev. Lett. {\bf 76}, 835 
(1996).
\bibitem{Chien} See, for e.g., T.R. Chien, T.W. Jing, N.P. Ong, and Z.Z. Wang, Phys. Rev. Lett. {\bf 66}, 
3075 (1991).
\bibitem{Xu} Z. A. Xu, N. P. Ong, Y. Wang, T. Kageshita, and S. Uchida, Nature {\bf 406},
    486 (2000).
\bibitem{Wang} Yayu Wang, Z. A. Xu, T. Kakeshita, S. Uchida, S. Ono, Yoichi Ando, and N. P. Ong, 
Phys. Rev. B {\bf 64}, 224519 (2001).
\bibitem{Corson} J. Corson, R. Mallozzi, J. Orenstein, J. N. Eckstein, and I. Bozovic,  Nature {\bf 398}, 
221 (1999).
\bibitem{Palstra}  T.T.M. Palstra, B. Batlogg, L. F. Schneemeyer, and J. V. Waszczak, Phys.
    Rev. Lett. {\bf 64}, 3090 (1990).
\bibitem{Ri} H. C. Ri, R. Gross, F. Gollnik, A. Beck, R. P. Huebener, P. Wagner, and H. Adrian, Phys. 
Rev. B {\bf 50}, 3312 (1994).
\bibitem{Clayhold} J. A. Clayhold, A. Linnen, F. Chen, and C. W. Chu, Phys. Rev. B {\bf 50}, 
    4252 (1994).
\bibitem{Capan} C. Capan {\em et al.} Phys. Rev. Lett. {\bf 88}, 056601 (2002).

\bibitem{Vidal} (a) Felix Vidal, Phys. Rev. B {\bf 8}, 1982 (1973); (b) O. L. de Lange and F. A. Otter, Jr. 
Jnl. Low Temp. Phys. {\bf 18}, 31 (1975); (c) R. P. Huebener and A. Seher, Phys. Rev. {\bf 181}, 701 
(1969). 
\bibitem{Onsager} The constitutive equations and Onsager reciprocity are discussed in Refs. 
\protect\cite{Maki}.
\bibitem{Maki} C. Caroli,  and K. Maki, Phys. Rev. {\bf 164}, 591 (1967); A. Houghton and K. Maki, 
Phys. Rev. B {\bf 3}, 1625 (1971); Chia-Ren Hu, Phys. Rev. B {\bf 13}, 4780 (1976).
\bibitem{fit}  Figure \ref{L20}c shows why previous attempts to fit Eq. \ref{Maki} to curves of 
$Ts_{\phi}$ vs. $T$ (taken at {\em fixed} $H<$15 T) were problematical.  The data lie outside the regime 
of the perturbation calculation.  
\bibitem{melting} Below 16 K in LSCO, $H_m(T)\sim {\rm e}^{-T/T_0}$ with $T_0\simeq$ 6 K.  Similar 
behavior of $H_m$ is seen in underdoped YBCO.  
\bibitem{Welp} U. Welp {\em et al.}, Phys. Rev. Lett. {\bf 62}, 1908 (1989).
\bibitem{Emery} V. J. Emery and S. A. Kivelson, Nature {\bf 374}, 434 (1995).

\bibitem{Wen} X. G. Wen, and P. A. Lee, Phys. Rev. Lett. {\bf 78},  4111 (1997); L. B. Ioffe and A.J. 
Millis, {\em preprint} 2002.
\bibitem{Pan} S. H. Pan, E. W. Hudson, A. K. Gupta, K. W. Ng, H. Eisaki, S. Uchida, and J. C. Davis, 
Phys. Rev. Lett. {\bf 85},  1536 (2000).
%
%

\end{references}
\end{document}